\documentstyle[12pt]{article}
\catcode `\@=11
\@addtoreset{equation}{section}

\newcommand{\be}{\begin{equation}}
\newcommand{\ee}{\end{equation}}
\newcommand{\R}{\rm I \mkern -3mu R}
\newcommand{\N}{\rm I \mkern -3mu N}
\newtheorem{thm}{Theorem}
\newtheorem{rem}{Remark}
\newtheorem{lemma}{Lemma}
\begin{document}
\begin{flushright}
IFA-FT-402-1994, November
\end{flushright}
\bigskip\bigskip\begin{center}
{\bf \Large{VARIATIONAL EQUATIONS AND\\
{}~\\SYMMETRIES IN THE LAGRANGIAN FORMALISM}}
\end{center}
\vskip 1.0truecm
\centerline{\bf
D. R. Grigore\footnote{e-mail: grigore@roifa.bitnet, grigore@ifa.ro}}
\vskip5mm
\centerline{Dept. Theor. Phys., Inst. Atomic Phys.,}
\centerline{Bucharest-M\u agurele, P. O. Box MG 6, ROM\^ANIA}
\vskip 2cm
\bigskip \nopagebreak \begin{abstract}
\noindent
Symmetries in the Lagrangian formalism of arbitrary order are
analysed with the help of the so-called Anderson-Duchamp-Krupka
equations. For the case of second order equations and a scalar
field we~establish a polynomial structure in the second order
derivatives. This structure can be used to make more precise the
form of a general symmetry. As an illustration we analyse the
case of Lagrangian equations with Poincar\'e invariance or with
universal invariance.
\end{abstract}

\section{Introduction}

The study of classical field theory in the framework of the
Lagrangian formalism is still a subject of active research. For
first-order Lagrangian systems one usually prefers the use of
the Poincar\'e-Cartan form or related geometrical objects (see
for instance \cite{Kl},\cite{G1}). For higher-order Lagrangian
systems it is difficult to find a proper generalization of the
Poincar\'e-Cartan form having the same properties as for the
first-order case. Particulary difficult is to find such a
generalization having a nice behaviour with respect to the
(Noetherian) symmetries. A way out is to use a related formalism
based on the Euler-Lagrange operator and its intrinsec
characterization by Helmholtz equations. In fact, it was noticed
sometimes ago that, in the case of second order differential
equations  describing a system with finite number of degrees of
freedom, one can give necessary and sufficient conditions such that the
equations follow from a Lagrangian: they are the so-called Helmholtz
equations (see \cite{OK} for a rather complete bibliography on this
problem). Remarcably, this result can be extended to the general
case of classical field theory and to equations of arbitrary order,
leading to the so-called Anderson-Duchamp-Krupka (ADK) equations
\cite{AD}, \cite{Kr}, which semms to be less known in the physics
literature. The proper framework for this formalism is based on the
jet-bundle structures.

The purpose of this paper is to prove that this formalism based
on the ADK equations can be used to treat rather completely
higher-order Lagrangian systems with groups of symmetries.

Section 2 has the purpose of presenting the formalism. For the
sake of the completeness we will also sketch the derivations of
the ADK equations. Section 3 is dedicated to the extensive study
of second-order Lagrangian equations. In the case of a scalar
field one can practically "solve" the ADK equations establishing
a polynomial structure in the second-order derivatives. This
central result greatly simplifies the study of (Noetherian)
symmetries.

In Section 4 we impose, in addition, invariance with respect to
some symmetry group. Combining with the result of Section 3 one
can completely analyse some interesting symmetry groups as the
Poincar\'e invariance and the so-called universal invariance \cite{F}.
Section 5 is dedicated to some final comments.

\section{A Higer-Order Lagrangian Formalism}

2.1 The kinematical structure of classical field theory is based
on a fibered bundle structure
$
\pi: S \mapsto M
$
where
$S$
and
$M$
are differentiable manifolds of dimensions
$
dim(M) = n,~dim(S) = N + n
$
and
$\pi$
is the canonical projection of the fibration. Usually
$M$
is interpreted as the "space-time" manifold and the fibers of
$S$
as the field variables. Next, one considers the
$k$-jet
bundle
$
J^{k}_{n}(S) \mapsto M~(k=0,...,p).
$
By convention
$
J^{0}_{n}(S) \equiv S
$
and
$
p \in \N \cup \{\infty\}.
$

One usually must take
$
p \in \N
$
but sufficienty large such that all formulas make sense.
Let us consider a local system of coordinates in the chart
$
U \subseteq S:~
(x^{\mu})~(\mu = 1,...,n).
$

Then on some chart
$
V \subseteq \pi^{-1}(U) \subset S
$
we take a local coordinate system adapted to the fibration structure:
$
(x^{\mu},\psi^{A})~(\mu = 1,...,n,~A = 1,...,N)
$
such that the canonical projection is
$
\pi(x^{\mu},\psi^{A}) = (x^{\mu}).
$

Then one can extend this system of coordinates to
$
J^{k}_{n}(S)
$
for any
$k \leq p$:
$$
(x^{\mu},\psi^{A},\psi^{A}_{\mu},...,\psi^{A}_{\mu_{1},...,\mu_{k}}),
{}~1 \leq \mu_{1} \leq \cdots \mu_{k} \leq n.
$$
If
$
\mu_{1},...,\mu_{k}
$
are arbitrary, then by
$
\{\mu_{1},...,\mu_{k}\}
$
we understand the operation of increasing ordering; then the notation
$
\psi^{A}_{\{\mu_{1},...,\mu_{k}\}}
$
makes sense obviously.

2.2 Let us consider
$
s < p
$
and
$T$
a
$(n + 1)$-form
which can be written in the local coordinates introduced above as:
\be
T = {\cal T}_{A}~d\psi^{A} \wedge dx^{1} \wedge \cdots \wedge dx^{n}
\label{edif}
\ee
with
$
{\cal T}_{A}
$
some smooth functions of
$
(x^{\mu},\psi^{A},\psi^{A}_{\mu},...,\psi^{A}_{\mu_{1},...,\mu_{s}}).
$

Then
$T$
can be globally defined. Indeed, if we make a change of charts
adapted to the fiber bundle structure:
\be
\phi(x^{\mu},\psi^{A}) = (f^{\mu}(x),F^{A}(x,\psi)) \
\label{chcharts1}
\ee
then in the new coordinates
$T$
has the same structure (\ref{edif}) as above. In fact, one
immediately gets that:
\be
{\cal T}_{A}' = det\left({\partial f^{\mu}\over \partial
x^{\nu}}\right) {\partial F^{B}\over \partial \psi^{A}}
{\cal T}_{B} \circ\dot\phi
\label{chcharts2}
\ee
where
$
\dot\phi
$
is the lift of
$
\phi
$
to
$
J^{k}_{n}(S).
$

We call such a
$T$
a {\it differential equation of order s}.

2.3 To introduce some special type of differential equations we
need some very useful notations \cite{AD}. We define the
differential operators:
\be
\partial^{\mu_{1},...,\mu_{l}}_{A} \equiv {r_{1}!...r_{l}! \over
l!} {\partial \over \partial \psi^{A}_{\{\mu_{1},...,\mu_{l}\}}}
\label{pdif}
\ee
for any
$
l = 0,...,k.
$
Here
$
r_{i}
$
is the number of times the index
$i$
appears in the sequence
$
\mu_{1},...,\mu_{l}.
$
The combinatorial factor in (\ref{pdif}) avoids possible
overcounting in the computations which will appear in the
following. One has then:
$$
\partial^{\mu_{1},...,\mu_{l}}_{A}
\psi^{B}_{\nu_{1},...,\nu_{l}} = {1\over l!} \delta^{A}_{B}
perm\left(\delta^{\mu_{i}}_{\nu_{j}}\right)~(\forall l \geq 0)
$$
and
$$
\partial^{\mu_{1},...,\mu_{l}}_{A}
\psi^{B}_{\nu_{1},...,\nu_{m}} = 0~~(l \not= m)
$$
where by
$
perm(A)
$
we mean the permanent of the matrix
$A$.

Next, we define the total derivative operators:
\be
D_{\mu} = {\partial\over \partial x^{\mu}} + \sum_{l \geq 0}
\psi^{A}_{\nu_{1},...,\nu_{l}\mu} \partial^{\nu_{1},...,\nu_{l}}_{A}.
\label{tdif}
\ee

One can check that
\be
D_{\mu}\psi^{A}_{\nu_{1},...,\nu_{l}} = \psi^{A}_{\nu_{1},...,\nu_{l}\mu}
\label{der}
\ee
\be
[D_{\mu}, D_{\nu}] = 0.
\label{com}
\ee

Finally we define the differential operators
\be
D_{\mu_{1},...,\mu_{l}} \equiv D_{\mu_{1}}...D_{\mu_{l}}.
\label{tdifs}
\ee

Because of (\ref{com}) the order of the factors in the right
hand side is irrelevant.

2.4 A differential equation
$T$
is called {\it locally variational} (or of the {\it
Euler-Lagrange type}) {\it iff} there exists a local real
function
${\cal L}$
such that the functions
$
{\cal T}_{A}
$
from (\ref{edif}) are of the form:
\be
{\cal E}_{A}({\cal L}) \equiv \sum_{l \geq 0} (-1)^{l}
D_{\mu_{1},...,\mu_{l}} (\partial^{\mu_{1},...,\mu_{l}}_{A}
{\cal L})
\label{Eop}
\ee

One calls
${\cal L}$
a {\it local Lagrangian} and:
\be
L \equiv {\cal L}~dx^{1}\wedge\cdots dx^{n}
\label{Lform}
\ee
a {\it local Lagrange form}. Let us note that
$L$
can be globally defined if we admit that at the change of charts
(\ref{chcharts1})
${\cal L}$
changes as follows:
\be
{\cal L}' =  det\left({\partial f^{\mu}\over \partial
x^{\nu}}\right)  {\cal L}\circ\dot\phi.
\label{chcharts3}
\ee

If the differential equation
$T$
is constructed as above then we denote it by
$
E(L).
$
A local Lagrangian is called a {\it total divergence} if it is of
the form:
\be
{\cal L} = D_{\mu} V^{\mu}.
\ee

One can check that in this case we have:
\be
E(L) = 0.
\label{trEL}
\ee

This property follows easily from:
\be
\left[\partial^{\mu_{1},...,\mu_{l}}_{A},D_{\nu}\right] = {1\over l}
\sum_{i=1}^{l} \delta_{\nu}^{\mu_{i}}
\partial^{\mu_{1},...,\hat{\mu_{i}},...,\mu_{l}}_{A},~(\forall l
\geq 0).
\ee

The converse of this statement is true if one works on
$
J_{n}^{\infty}(S)
$
(see \cite{T}). It is not known if this is true on
$
J_{n}^{p}(S)
$
with
$p$
finite. A local Lagrangian verifying (\ref{trEL}) is called {\it
trivial}.

2.5 Now we come to the central result from \cite{AD}, \cite{Kr}.
\begin{thm}
Let
$T$
be a differential equation of order
$s$. Then
$T$
is locally variational {\it iff} the functions
$
{\cal T}_{A}
$
from (\ref{edif}) verify the following equations:
\be
\partial^{\mu_{1},...,\mu_{l}}_{A} {\cal T}_{B} = \sum_{p=l}^{s}
(-1)^{p} C_{p}^{l} D_{\mu_{l+1},...,\mu_{p}}
\partial^{\mu_{1},...,\mu_{p}}_{B} {\cal T}_{A},~(l = 0,...,s).
\label{ADK}
\ee
\label{AD}
\end{thm}

\begin{rem}
These are the so-called {\it Anderson-Duchamp-Krupka
equations}. For
$
n = 1
$
and
$
s = 2
$
one obtains the well-known Helmholtz equations.
\end{rem}

{\bf Sketch of the proof} \cite{AD}

We remind the reader that we are working on
$
J^{p}_{n}(S)
$
with
$p$
sufficiently large.

$\Longrightarrow$ :

Suppose that
${\cal L}$
is a local Lagrangian depending on
$
(x^{\mu},\psi^{A},\psi^{A}_{\mu},...,\psi^{A}_{\mu_{1},...,\mu_{r}})
$
with
$
2r \geq s.
$

Then we must show that
$
{\cal T}_{A} = {\cal E}_{A}(L)
$
verify the ADK equations. The idea is the following: let
$
y^{A}~(A = 1,...,N)
$
some
$x$-dependent functions and:
$$
y^{A}_{\mu_{1},...,\mu_{l}} \equiv {\partial^{l} y^{A}\over
\partial x^{\mu_{1}}... \partial x^{\mu_{l}}}~~(\forall l = 0,...,2r).
$$

We define (locally) the vector field
$Y$
by:
$$
Y = \sum_{i=0}^{2r} y^{A}_{\mu_{1},...,\mu_{l}}
\partial^{\mu_{1},...,\mu_{l}}_{A}.
$$

One proves by direct computations that:
$$
{\cal L}_{Y}(L) = i_{Y} E(L) + L_{0}.
$$

Here
$
{\cal L}_{Y}
$
and
$
i_{Y}
$
are the standard operations of Lie derivative and inner
contraction.
$
L_{0}
$
is a Lagrange form corresponding to the trivial Lagrangian
$
{\cal L}_{0} = D_{\mu} V^{\mu}
$
where:
$$
V^{\mu} \equiv \sum_{p=1}^{r} \sum_{l=1}^{p} (-1)^{l+1}
y^{A}_{\mu_{l+1},...,\mu_{p}} D_{\mu_{1},...,\mu_{l-1}}
\left(\partial^{\mu_{1},...,\hat{\mu_{l}},...,\mu_{p}\nu}_{A}
{\cal L}\right).
$$

So one has evidently:
$$
E({\cal L}_{Y}(L)) = E(i_{Y} E(L)).
$$

But the Euler-Lagrange operator
$E$
contains only the operators
$
\partial^{\mu_{1},...,\mu_{l}}_{A}
$
and
$
D_{\mu}
$
(see (\ref{Eop})) and one can check directly that both commute
with
$
{\cal L}_{Y}
$
when applied to
$L;$
so
$E$
commutes with
$
{\cal L}_{Y}
$
and the preceeding relation implies:
$$
{\cal L}_{Y}(T) = E(i_{Y}T).
$$

The ADK equations are nothing but the coefficients of
$
y^{A}_{\mu_{1},...,\mu_{l}}
$
in this equation.

$\Longleftarrow$ :

Suppose that the differential equation
$T$
verifies (locally) the equations (\ref{ADK}). One can choose the
system of local coordinates such that
$
{\cal T}_{A}
$
are regular functions in the point:
$
\psi^{A}_{\mu_{1},...,\mu_{l}} = 0~(l = 0,...,s).
$
Then one defines the (local) Lagrangian:
\be
{\cal L} = \int_{0}^{1} \psi^{A} {\cal T}_{A}\circ \chi_{\lambda}
d\lambda
\label{Tonti}
\ee
where
$$
\chi_{\lambda} (x^{\mu},\psi^{A},\psi^{A}_{\mu},
...,\psi^{A}_{\mu_{1},...,\mu_{p}}) =
(x^{\mu},\lambda\psi^{A},\lambda\psi^{A}_{\mu},
...,\lambda\psi^{A}_{\mu_{1},...,\mu_{p}}).
$$

Then by direct computations one gets that
$
{\cal T}_{A} = {\cal E}_{A}(L).
$
$\Box$

The expression (\ref{Tonti}) is called the {\it Tonti Lagrangian}.

2.6 One would like to show that the ADK equations have a global
meaning i.e. if in some chart
$
{\cal T}_{A}
$
verify (\ref{ADK}), then
$
{\cal T}_{A}'
$
given by (\ref{chcharts2}) verify (\ref{ADK}). Suppose that
$
{\cal T}_{A}
$
verify (\ref{ADK}). Then Theorem 1 shows that
$
{\cal T}_{A} = {\cal E}_{A}(L)
$
for some Lagrange form
$L$.
If we consider a change of coordinates
$\phi$
on
$S$
(see subsection 2.2) one can prove that
\be
{\cal T}_{A}' = {\cal E}(L')
\label{chcharts4}
\ee
where
$L'$
is the Lagrange form associated to the Lagrangian given by
(\ref{chcharts3}). We apply again Theorem 1 and obtain that
$
{\cal T}_{A}'
$
verify again (\ref{ADK}). Let us give the idea of the proof of
(\ref{chcharts4}). An {\it evolution} is any section
$
\Psi:M \rightarrow S
$
of the bundle
$
\pi: S \mapsto M.
$

Let us denote by
$
\dot\Psi:M \rightarrow J^{s}_{n}(S)
$
the natural lift of
$
\Psi
$
and define the action functional by:
\be
{\cal A}_{L}(\Psi) \equiv \int (\dot\Psi)^{*} L.
\ee

The fundamental formula of the variational calculus is then:
\be
\delta_{X}{\cal A}_{L}(\Psi) \equiv \int (\dot\Psi)^{*} i_{X}E(L)
\ee
where
$
X \equiv X^{A} {\partial \over \partial \psi^{A}}
$
is the infinitesimal variation. One computes in two obvious ways
this variation and discovers that
$
{\cal E}_{A}(L)
$
and
$
{\cal E}_{A}(L')
$
are connected by a relation of the type (\ref{chcharts2}). From
this the equation (\ref{chcharts4}) follows immediately.

2.7 Let us suppose that
$T$
is a differential equation and
$
\pi: S \mapsto M
$
is a evolution. One says that
$
\Psi
$
is a {\it solution} of
$T$
if one has:
\be
(\dot\Psi)^{*}~T = 0.
\label{ELeq}
\ee

If
$T$
is locally variational
$
T = E(L)
$
one obtains the global form of the {\it Euler-Lagrange equations}.
In local coordinates one can arrange such that
$
\Psi
$
has the form
$
x^{\mu} \mapsto (x^{\mu},\Psi(x));
$
then
$
\dot\Psi:M \rightarrow J^{s}_{n}(S)
$
is given by
$$
x^{\mu} \mapsto \left(x^{\mu},\Psi(x),
{\partial \Psi\over \partial x^{\mu}}(x),
...,{\partial \Psi\over \partial x^{\mu_{1}}...\partial x^{\mu_{s}}}(x)
\right)
$$
and (\ref{ELeq}) take the well-known form.

2.8 We come now to the notion of symmetry. By a {\it symmetry} of
$T$
we understand a map
$
\phi \in Diff(S)
$
such that if
$
\Psi:M \rightarrow S
$
is a solution of
$T$,
then
$
\phi \circ \Psi
$
is a solution of
$T$
also.

It is tempting to try to classify all possible symmetries
associated to a given
$T.$
In general, this problem is too difficult to tackle. We will solve
a particular case in the next section. For the moment we content
ourselves to note that if
$
\phi
$
verifies:
\be
(\dot\phi)^{*}~T = \lambda T,~(\lambda \in \R^{*})
\ee
$
(\dot\phi \in Diff(J_{n}^{s}(S))
$
being the natural lift of
$
\phi
$),
then
$
\phi
$
is a symmetry. For
$
\lambda = 1
$
these are the so-called {\it Noetherian symmetries}. Indeed if
$
T = E(L)
$
one can recover the usual definition:
\be
{\cal A}_{L}(\phi\circ\Psi) = {\cal A}_{L}(\Psi) + a~trivial~action
\label{sim2}
\ee
where by a {\it trivial action} we mean an action
$
{\cal A}_{L_{0}}
$
with
$
L_{0}
$
a trivial Lagrangian. Noetherian symmetries are important
because from (\ref{sim2}) one can obtain conservation laws.

\section{Second Order Euler-Lagrange Equations}

3.1 We particularize the ADK equations for case
$
s = 2
$
of second-order Euler-Lagrange equations. It is not hard to obtain
the following set of equations:
\be
\partial_{A}^{\mu_{1}\mu_{2}} {\cal T}_{B} =
\partial_{B}^{\mu_{1}\mu_{2}} {\cal T}_{A}
\label{ADK1}
\ee
\be
\left(\partial_{B}^{\mu\rho_{1}}\partial_{C}^{\rho_{2}\rho_{3}} +
\partial_{B}^{\mu\rho_{2}}\partial_{C}^{\rho_{3}\rho_{1}} +
\partial_{B}^{\mu\rho_{3}}\partial_{C}^{\rho_{1}\rho_{2}}\right)
{\cal T}_{A} = 0
\label{ADK2}
\ee
\be
\partial_{A}^{\mu_{1}} {\cal T}_{B} +
\partial_{B}^{\mu_{1}} {\cal T}_{A} =
2\left(\partial_{\mu_{2}} + \psi^{C}_{\mu_{2}} \partial_{C} +
\psi^{C}_{\{\mu_{2}\nu_{1}\}} \partial_{C}^{\nu_{1}}\right)
\partial_{B}^{\mu_{1}\mu_{2}}{\cal T}_{A}
\label{ADK3}
\ee
\begin{eqnarray}
&~&\partial_{A} {\cal T}_{B} - \partial_{B} {\cal T}_{A} =
-\left(\partial_{\mu_{1}} + \psi^{C}_{\mu_{1}} \partial_{C} +
\psi^{C}_{\{\mu_{1}\nu_{1}\}} \partial_{C}^{\nu_{1}}\right)
\partial_{B}^{\mu_{1}}{\cal T}_{A} + \nonumber
\\
&~&\left(\partial_{\mu_{1}} + \psi^{C}_{\mu_{1}} \partial_{C} +
\psi^{C}_{\{\mu_{1}\nu_{1}\}} \partial_{C}^{\nu_{1}}\right)
\left(\partial_{\mu_{2}} + \psi^{D}_{\mu_{2}} \partial_{D} +
\psi^{D}_{\{\mu_{2}\nu_{2}\}} \partial_{D}^{\nu_{2}}\right)
\partial_{B}^{\mu_{1}\mu_{2}}{\cal T}_{A}.
\label{ADK4}
\end{eqnarray}

It is plausible to conjecture that from (\ref{ADK1}) and
(\ref{ADK2}) follows that
$
{\cal T}_{A}
$
is a polynomial in the second order derivatives
$
\psi^{A}_{\{\mu\nu\}}.
$
We have succeded to prove this conjecture for the case of the
scalar field
$
N = 1.
$

3.2 Let
$
M \simeq \R^{n}
$
with coordinates
$
(x^{\mu})~\mu = 1,...,n
$
and
$
S \subset M \times \R
$
with  coordinates
$
(x^{\mu},\psi).
$
We can imbed naturally
$
J^{k}_{n}(S)
$
in an Euclidean space with coordinates
$
(x^{\mu},\psi,\psi_{\mu},...,\psi_{\mu_{1},...,\mu_{k}}).
$

Then we have from (\ref{edif}), (\ref{pdif}) and (\ref{tdif}):
\be
T = {\cal T}~d\psi \wedge dx^{1} \wedge \cdots \wedge dx^{n}
\ee
\be
\partial^{\mu_{1},...,\mu_{l}} \equiv {r_{1}!...r_{l}! \over
l!} {\partial \over \partial \psi_{\{\mu_{1},...,\mu_{l}\}}}
\ee
\be
D_{\mu} = {\partial \over\partial x^{\mu}} + \sum_{l \geq 0}
\psi_{\{\nu_{1},...,\nu_{l}\mu\}} \partial^{\nu_{1},...,\nu_{l}}.
\ee

The ADK equations (\ref{ADK1})-(\ref{ADK4}) simplify
considerably. In fact (\ref{ADK1}) is trivial, (\ref{ADK2}) and
(\ref{ADK3}) become
\be
\left(\partial^{\mu\rho_{1}}\partial^{\rho_{2}\rho_{3}} +
\partial^{\mu\rho_{2}}\partial^{\rho_{3}\rho_{1}} +
\partial^{\mu\rho_{3}}\partial^{\rho_{1}\rho_{2}}\right)
{\cal T}
\label{ADK'}
\ee
and respectively
\be
\partial^{\mu} {\cal T} = \left({\partial\over\partial x^{\nu}} +
\psi_{\nu} \partial + \psi_{\{\nu\rho\}} \partial^{\rho}\right)
\partial^{\mu\nu}{\cal T}.
\label{ADK''}
\ee

Finally (\ref{ADK4}) is a consequence of (\ref{ADK''}). We will
be able to prove that (\ref{ADK'}) is compatible with a
polynomial structure of
$T$
in the second order derivatives
$
\psi_{\{\mu\nu\}}.
$

3.3 Let us note that Tonti Lagrangian associated to
$T$
is (see (\ref{Tonti})):
\be
{\cal L} = \int_{0}^{1} \psi {\cal T}\circ \chi_{\lambda}
d\lambda
\ee
with
$$
\chi_{\lambda} (x^{\mu},\psi,\psi_{\mu},\psi_{\{\mu\nu\}}) =
(x^{\mu},\lambda\psi,\lambda\psi_{\mu},\lambda\psi_{\{\mu\nu\}}).
$$

The Euler-Lagrange equations for
${\cal L}$
are {\it a priori} of order fourth because the Lagrangian is of
second order (see (\ref{Eop})). It follows that there are some
constraints on
${\cal L}$;
namely one should require that the terms of third and fourth
order in the expression
$
E(L)
$
should be identically zero. It is easy to prove that this
condition amounts to:
\be
\left(\partial^{\mu\rho_{1}}\partial^{\rho_{2}\rho_{3}} +
\partial^{\mu\rho_{2}}\partial^{\rho_{3}\rho_{1}} +
\partial^{\mu\rho_{3}}\partial^{\rho_{1}\rho_{2}}\right)
{\cal L} = 0
\label{constr}
\ee

More precisely we have

\begin{lemma}
A second order Lagrangian
${\cal L}$
leads to second order Euler-Lagrange equations if and only if it
verifies the relation (\ref{constr}). In this case we have:
\begin{eqnarray}
&{\cal E}(L)& = \partial {\cal L}
-\left({\partial\over \partial x^{\mu}} + \psi_{\mu} \partial +
\psi_{\{\mu\nu\}} \partial^{\nu}\right) \partial^{\mu}{\cal L}
\nonumber \\
&+& \left({\partial\over \partial x^{\mu_{1}}} +
\psi_{\mu_{1}} \partial + \psi_{\{\mu_{1}\nu_{1}\}} \partial^{\nu_{1}}\right)
\left({\partial\over \partial x^{\mu_{2}}} + \psi_{\mu_{2}} \partial +
\psi_{\{\mu_{2}\nu_{2}\}} \partial^{\nu_{2}}\right)
\partial^{\mu_{1}\mu_{2}}{\cal L}\nonumber \\
{}~
\end{eqnarray}
\end{lemma}

3.4 We turn now to the study of the equations (\ref{ADK'}) (or
(\ref{constr})). Let us define the expressions:
\be
\psi^{\mu_{1},...,\mu_{k};\nu_{1},...,\nu_{k}} \equiv
{1\over (n-k)!} \varepsilon^{\mu_{1},...,\mu_{n}}
\varepsilon^{\nu_{1},...,\nu_{n}} \prod_{i=k+1}^{n}
\psi_{\{\mu_{i}\nu_{i}\}}~(\forall k = 0,...,n)
\label{minors}
\ee

Up to a sign,
$
\psi^{\mu_{1},...,\mu_{k};\nu_{1},...,\nu_{k}}
$
is the determinant of the matrix
$
\psi_{\{\mu\nu\}}
$
with the lines
$
\mu_{1},...,\mu_{k}
$
and the columns
$
\nu_{1},...,\nu_{k}
$
deleted. The combinatorial factor is chosen such that:
\be
\psi^{\emptyset,\emptyset} = det(\psi_{\{\mu\nu\}}).
\ee

We prove now

\begin{thm}
The general solution of the equations (\ref{ADK'}) is of the
following form:
\be
{\cal T} = \sum_{k=0}^{n} {1\over (k!)^{2}}
{\cal T}_{\mu_{1},...,\mu_{k};\nu_{1},...,\nu_{k}}
\psi^{\mu_{1},...,\mu_{k};\nu_{1},...,\nu_{k}}
\label{polyn}
\ee
where
$
{\cal T}_{...}
$
are independent of
$
\psi_{\{\mu\nu\}}
$:
\be
\partial^{\rho\sigma} {\cal
T}_{\mu_{1},...,\mu_{k};\nu_{1},...,\nu_{k}} = 0,~(\forall k = 0,...,n)
\ee
and have the same symmetry properties as
$
\psi^{...}
$:
complete antisymmetry in
$
\mu_{1},...,\mu_{k},
$
complete antisymmetry in
$
\nu_{1},...,\nu_{k}
$
and symmetry with respect to the interchange:
$
\mu_{1},...,\mu_{k} \leftrightarrow \nu_{1},...,\nu_{k}.
$
\label{structure}
\end{thm}

{\bf Proof}

(i) One uses induction over
$n$.
For
$
n = 2,
$
the equations (\ref{ADK'}) are simple to write and one obtains
indeed that the general solution is of the form (\ref{polyn}).
We suppose that we have the assertion of the theorem for a given
$n$
and we prove it for
$
n + 1.
$
In this case the indices
$
\mu,\nu, ...
$
takes values (for notational convenience)
$
\mu,\nu, ...= 0,...,n
$
and
$
i,j,...= 1,...,n.
$
If we consider in (\ref{ADK'}) that
$
\mu,\rho_{1},\rho_{2},\rho_{3} = 1,...,n
$
then we can apply the induction hypotesis and we get.
\be
{\cal T} = \sum_{k=0}^{n} {1\over (k!)^{2}}
\tilde{\cal T}_{i_{1},...,i_{k};j_{1},...,j_{k}}
\tilde{\psi}^{i_{1},...,i_{k};j_{1},...,j_{k}}
\label{polyn'}
\ee

Here
$
\tilde{\cal T}
$
has obvious symmetry properties and can depend on
$
x, \psi, \psi_{\mu}
$
{\it and}
$
\psi_{\{0\mu\}}.
$
The minors
$
\tilde{\psi}^{...}
$
are constructed from the matrix
$
\psi_{\{ij\}}
$
according to the prescription (\ref{minors}).

(ii) We still have at our disposal the relation (\ref{ADK'})
where at least one index takes the value
$0$.
We obtain rather easily:
\be
\left(\partial^{00}\right)^{2} \tilde{\cal
T}_{i_{1},...,i_{k};j_{1},...,j_{k}}  = 0,~(\forall k = 0,...,n)
\label{eq1}
\ee
\be
\partial^{00}\partial^{0l}  \tilde{\cal
T}_{i_{1},...,i_{k};j_{1},...,j_{k}}  = 0,~(\forall k = 0,...,n)
\label{eq2}
\ee
\be
\partial^{0l}\partial^{0m} \tilde{\cal T}_{\emptyset;\emptyset} = 0
\label{eq3}
\ee
\begin{eqnarray}
&{1\over 2}& \sum_{p,q=1}^{k} (-1)^{p+q}
\left(\delta_{i_{p}}^{m}\delta_{j_{q}}^{l} + \delta_{i_{p}}^{l}
\delta_{j_{q}}^{m}\right) \partial^{00} \tilde{\cal
T}_{i_{1},...,\hat{i_{p}},...i_{k};j_{1},...,\hat{j_{q}},...,j_{k}} +
\nonumber \\
&2& \partial^{0l} \partial^{0m}\tilde{\cal
T}_{i_{1},...,i_{k};j_{1},...,j_{k}} = 0,
{}~~(\forall k = 1,...,n)
\label{eq4}
\end{eqnarray}
\be
\sum_{(l,m,r)} \sum_{p,q=1}^{k} (-1)^{p+q}
\left(\delta_{i_{p}}^{m}\delta_{j_{q}}^{r} + \delta_{i_{p}}^{r}
\delta_{j_{q}}^{m}\right) \partial^{0l} \tilde{\cal
T}_{i_{1},...,\hat{i_{p}},...i_{k};j_{1},...,\hat{j_{q}},...,j_{k}}  = 0~~
(\forall k = 1,...,n).
\label{eq5}
\ee

Here by
$
\sum_{(l,m,r)}
$
we understand the sum over all cyclic permutations of the indices
$
l,m,r.
$

It is a remarcable fact that these equations can be solved i.e.
one can describe the most general solution.

{}From (\ref{eq1}) we have:
\be
\tilde{\cal T}_{i_{1},...,i_{k};j_{1},...,j_{k}} =
{\cal T}^{(0)}_{i_{1},...,i_{k};j_{1},...,j_{k}} +
\psi_{\{00\}} {\cal T}^{(1)}_{i_{1},...,i_{k};j_{1},...,j_{k}}~
(k = 0,...,n)
\label{sol}
\ee
with the restrictions:
\be
\partial^{00} {\cal T}^{(l)}_{i_{1},...,i_{k};j_{1},...,j_{k}} = 0,~
(k = 0,...,n;~l = 0,1)
\label{restr1}
\ee

{}From (\ref{eq2}) and (\ref{eq3}) we also get:
\be
\partial^{0l} {\cal T}^{(1)}_{i_{1},...,i_{k};j_{1},...,j_{k}} = 0,~
(k = 0,...,n)
\ee
\be
\partial^{0l}\partial^{0m} {\cal T}^{(0)}_{\emptyset;\emptyset} = 0,~
\label{restr3}
\ee

Finally (\ref{eq4}) and (\ref{eq5}) become:
\begin{eqnarray}
&{1\over 2}& \sum_{p,q=1}^{k} (-1)^{p+q}
\left(\delta_{i_{p}}^{m}\delta_{j_{q}}^{l} + \delta_{i_{p}}^{l}
\delta_{j_{q}}^{m}\right) \partial^{00} {\cal
T}^{(1)}_{i_{1},...,\hat{i_{p}},...i_{k};j_{1},...,\hat{j_{q}},...,j_{k}} +
\nonumber \\
&2& \partial^{0l} \partial^{0m}{\cal
T}^{(0)}_{i_{1},...,i_{k};j_{1},...,j_{k}} = 0,~(\forall k = 1,...,n)
\label{eq4'}
\end{eqnarray}
\be
\sum_{(l,m,r)} \sum_{p,q=1}^{k} (-1)^{p+q}
\left(\delta_{i_{p}}^{m}\delta_{j_{q}}^{r} + \delta_{i_{p}}^{r}
\delta_{j_{q}}^{m}\right) \partial^{0l} {\cal
T}^{(0)}_{i_{1},...,\hat{i_{p}},...i_{k};j_{1},...,\hat{j_{q}},...,j_{k}}  = 0~
(\forall k = 1,...,n)
\label{eq5'}
\ee

(iii) Now the analysis can be pushed further if we apply the operator
$
\partial^{0r}
$
to (\ref{eq4'}); taking into account (\ref{restr1}) we obtain:
\be
\partial^{0r}\partial^{0l} \partial^{0m} {\cal
T}^{(0)}_{i_{1},...,i_{k};j_{1},...,j_{k}} = 0~(\forall k = 1,...,n)
\label{eq4''}
\ee

The equations (\ref{restr3}) and (\ref{eq4''}) can be used to
obtain rather easily a polynomial structure in
$
\psi_{\{0l\}}.
$
The details are elementary and one gets from (\ref{restr3}):
\be
{\cal T}^{(0)}_{\emptyset;\emptyset} =
{\cal T}^{\emptyset}_{\emptyset;\emptyset} +
\sum_{l} {\cal T}^{l}_{\emptyset;\emptyset} \psi_{\{0l\}}
\label{sol1}
\ee
with
\be
\partial^{0\mu} {\cal T}^{...}_{\emptyset;\emptyset} = 0.
\label{restr3'}
\ee

Analogously, one establishes from (\ref{eq4''}):
\begin{eqnarray}
{\cal T}^{(0)}_{i_{1},...,i_{k};j_{1},...,j_{k}}& =&
{\cal T}^{\emptyset}_{i_{1},...,i_{k};j_{1},...,j_{k}} +
\sum_{l} {\cal T}^{l}_{i_{1},...,i_{k};j_{1},...,j_{k}} \psi_{\{0l\}} +
\nonumber \\
&{1\over 2}& \sum_{l,m} {\cal T}^{lm}_{i_{1},...,i_{k};j_{1},...,j_{k}}
\psi_{\{0l\}}\psi_{\{0m\}},~(\forall k = 1,...,n).
\label{sol2}
\end{eqnarray}

Here we have:
\be
\partial^{0\mu} {\cal T}^{...}_{i_{1},...,i_{k};j_{1},...,j_{k}} = 0~
(k = 0,...,n).
\label{restr4}
\ee

We can also suppose that:
\be
{\cal T}^{lm}_{i_{1},...,i_{k};j_{1},...,j_{k}} =~
{\cal T}^{ml}_{i_{1},...,i_{k};j_{1},...,j_{k}}.
\ee

If we insert (\ref{sol2}) into (\ref{eq4'}) we get:
\be
{\cal T}^{lm}_{i_{1},...,i_{k};j_{1},...,j_{k}} =
- {1\over 2} \sum_{p,q=1}^{k} (-1)^{p+q}
\left(\delta_{i_{p}}^{m}\delta_{j_{q}}^{l} + \delta_{i_{p}}^{l}
\delta_{j_{q}}^{m}\right) {\cal T}^{(1)}_
{i_{1},...,\hat{i_{p}},...i_{k};j_{1},...,\hat{j_{q}},...,j_{k}}.
\label{sol3}
\ee

Finally, inserting (\ref{sol2}) into (\ref{eq5'}) we get:
\be
\sum_{(l,m,r)} \sum_{p,q=1}^{k} (-1)^{p+q}
\left(\delta_{i_{p}}^{m}\delta_{j_{q}}^{r} + \delta_{i_{p}}^{r}
\delta_{j_{q}}^{m}\right) {\cal T}^{l}_
{i_{1},...,\hat{i_{p}},...i_{k};j_{1},...,\hat{j_{q}},...,j_{k}}  = 0~
(\forall k = 1,...,n)
\label{eq6}
\ee
\be
\sum_{(l,m,r)} \sum_{p,q=1}^{k} (-1)^{p+q}
\left(\delta_{i_{p}}^{m}\delta_{j_{q}}^{r} + \delta_{i_{p}}^{r}
\delta_{j_{q}}^{m}\right) {\cal T}^{ls}_
{i_{1},...,\hat{i_{p}},...i_{k};j_{1},...,\hat{j_{q}},...,j_{k}}  = 0~
(\forall k = 1,...,n)
\label{eq7}
\ee

Let us summarize what we have obtained up till now. The solution
of (\ref{eq1})-(\ref{eq5}) is given by (\ref{sol}) where
$
{\cal T}_{...}^{(0)}
$
is given by (\ref{sol2}) with
$
{\cal T}_{...}^{lm}
$
explicitated by (\ref{sol3}) and
$
{\cal T}_{...}^{l}
$
restricted by (\ref{eq6}). One also has to keep in mind
(\ref{restr3'}) and (\ref{restr4}). We will show that
(\ref{sol3}) identically verifies (\ref{eq7}) so in fact we are
left to solve only (\ref{eq6}).

(iv) It is rather strange that equations of the type (\ref{eq6})
and (\ref{eq7}) can be analysed using techniques characteristic
to quantum mechanics, namely the machinery of Fock space. In
fact, let us consider the antisymmetric Fock space
$
{\cal F}^{(-)}(\R^{n});
$
we define next the Hilbert space
$
{\cal H} \equiv {\cal F}^{(-)}(\R^{n}) \otimes {\cal
F}^{(-)}(\R^{n})
$

It is clear that the tensors
$
{\cal T}^{...}_{i_{1},...,i_{k};j_{1},...,j_{k}}
$
can be viewed as elements of
${\cal H}$
verifying also the symmetry property:
\be
S {\cal T}^{...} = {\cal T}^{...}
\label{sym}
\ee
where
\be
S(\phi\otimes\psi) = \psi\otimes\phi,~(\forall \phi,\psi \in
{\cal F}^{(-)}(\R^{n})).
\ee

Let us denote by
$
a_{(l)}
$
and
$
a^{*(l)},~(\forall l = 1,...,n)
$
the annihilation and respectively the creation operators acting in
$
{\cal F}^{(-)}(\R^{n});
$
then we have in
${\cal H}$
the operators:
$$
b^{*(l)} \equiv a^{*(l)}\otimes {\bf 1},~c^{*(l)} \equiv {\bf
1}\otimes a^{*(l)}
$$
and similarly for
$
b_{(l)}
$
and
$
c_{(l)}.
$
In these notations (\ref{sol3}) and (\ref{eq6}), (\ref{eq7})
become:
\be
{\cal T}^{lm} = - {1\over 2k^{2}} \left[b^{*(l)}c^{*(m)} +
b^{*(m)}c^{*(l)}\right] {\cal T}^{(1)}
\label{sol3'}
\ee
\be
\sum_{(l,m,r)} \left[b^{*(l)}c^{*(m)} + b^{*(m)}c^{*(l)}\right]
{\cal T}^{r} = 0
\label{eq6'}
\ee
\be
\sum_{(l,m,r)} \left[b^{*(l)}c^{*(m)} + b^{*(m)}c^{*(l)}\right]
{\cal T}^{rs} = 0
\label{eq7'}
\ee

Now it is extremely easy to prove that (\ref{sol3'}) identically
verifies (\ref{eq7'}) so, in fact, (\ref{sol3}) identically verifies
(\ref{eq7}) as we have announced above.

We concentrate now on (\ref{eq6'}). If we take
$
l = m = r
$
we get:
$$
b^{*(l)}c^{*(l)}{\cal T}^{l} = 0~(\textstyle{no~summation~over~l!})
$$
which easily implies that
$
{\cal T}^{l}
$
must have the following structure:
$$
{\cal T}^{l} = b^{*(l)} B + c^{*(l)} C + b^{*(l)} c^{*(l)} D
$$
with
$
B, C
$
and
$D$
obtained from the vacuum by  applying polynomial operators in
all creation operators with the exception of
$
b^{*(l)}
$
and
$
c^{*(l)}.
$

{}From (\ref{sym}) we get:
$$
C = SB,~D = SD
$$
so, in fact,
$
{\cal T}^{l}
$
is of the form:
\be
{\cal T}^{l} = b^{*(l)} B' + c^{*(l)} SB'
\label{sol4}
\ee
with
$
B'
$
arbitrary. Now it is easy to prove that (\ref{sol4}) identically
verifies (\ref{eq6'}) so it is the most general solution of this
equation.

Reverting to index notations, it follows that the most general
solution of (\ref{eq6}) is of the form:
\be
{\cal T}^{l}_{i_{1},...,i_{k};j_{1},...,j_{k}} =
\sum_{p=1}^{k} (-1)^{p-1} \left(\delta_{i_{p}}^{l} {\cal
T}_{i_{1},...,\hat{i_{p}},...i_{k};j_{1},...,j_{k}} +
\delta_{j_{p}}^{l} {\cal
T}_{j_{1},...,\hat{j_{p}},...,j_{k};i_{1},...,i_{k}}\right)
\label{sol5}
\ee
where
$
{\cal T}_{i_{2},...,i_{k};j_{1},...,j_{k}}
$
is completely antisymmetric in
$
i_{2},...,i_{k}
$
and in
$
j_{1},...,j_{k}.
$

The structure of
$
{\cal T}_{i_{1},...,i_{k};j_{1},...,j_{k}}
$
is completely elucidated: it is given by (\ref{sol}) where
$
{\cal T}^{(0)}_{...}
$
is given by (\ref{sol1}) and (\ref{sol2}); in (\ref{sol2})
$
{\cal T}^{l}_{...}
$
is given by (\ref{sol5}) and
$
{\cal T}^{lm}_{...}
$
by (\ref{sol3}). Everything depends on some arbitrary functions
$
{\cal T}^{...}_{\emptyset;\emptyset}
$
and
$
{\cal T}^{(1)}_{...}
$
which do not depend on
$
\psi_{\{0\mu\}}.
$

(v) It remains to introduce the expression for
$
\tilde{\cal T}
$
in (\ref{polyn'}) and regroup the terms. If we define:
$$
{\cal T}_{\emptyset;\emptyset} \equiv
{\cal T}^{\emptyset}_{\emptyset;\emptyset}
$$
$$
{\cal T}_{0,i_{1},...,i_{k};0j_{1},...,j_{k}} \equiv (k+1)
{\cal T}^{\emptyset}_{i_{1},...,i_{k};j_{1},...,j_{k}}
$$
$$
{\cal T}_{i_{1},...,i_{k};j_{1},...,j_{k}} \equiv
{\cal T}^{(1)}_{i_{1},...,i_{k};j_{1},...,j_{k}}
$$
$$
{\cal T}_{0,i_{2},...,i_{k};j_{1},...,j_{k}} \equiv
k {\cal T}_{i_{2},...,i_{k};j_{1},...,j_{k}}
$$
then (\ref{polyn'}) goes into (\ref{polyn}).
$\Box$

3.5 We can insert the solution (\ref{polyn}) of (\ref{ADK'})
into (\ref{ADK''}) and obtain some restrictions on the functions
$
{\cal T}_{...}.
$

It is convenient to define:
\be
{\delta\over\delta x^{\mu}} \equiv {\partial\over\partial
x^{\mu}} + \psi_{\mu} \partial
\ee

Then we obtain:
\begin{eqnarray}
&~&\sum_{i,j=1}^{k} (-1)^{i+j} \left( \delta_{\mu_{i}}^{\rho}
{\delta {\cal T}_{\mu_{1},...,\hat{\mu_{i}},...,\mu_{k};
\nu_{1},...,\hat{\nu_{j}},...,\nu_{k}}\over\delta x^{\nu_{j}}} +
\delta_{\nu_{j}}^{\rho}
{\delta {\cal T}_{\mu_{1},...,\hat{\mu_{i}},...,\mu_{k};
\nu_{1},...,\hat{\nu_{j}},...,\nu_{k}}\over\delta
x^{\mu_{i}}}\right) = \nonumber
\\
&~&\sum_{i=1}^{k} (-1)^{i-1} \left[ \delta_{\mu_{i}}^{\rho}
\left(\partial^{\lambda}
{\cal T}_{\lambda,\mu_{1},...,\hat{\mu_{i}},...,\mu_{k};
\nu_{1},...,\nu_{k}}\right) + \delta_{\nu_{i}}^{\rho} \left(
\partial^{\lambda} {\cal T}_{\mu_{1},...,\mu_{k};
\lambda,\nu_{1},...,\hat{\nu_{i}},...,\nu_{k}}\right)\right]
\label{ADK_1}
\end{eqnarray}
$
\hfill{(k = 1,...,n-1),}
$
\begin{eqnarray}
&~&\partial^{\rho}
{\cal T}_{\mu_{1},...,\mu_{n};\nu_{1},...,\nu_{n}} = \nonumber
\\
&~&{1\over2} \sum_{i,j=1}^{n} (-1)^{i+j} \left( \delta_{\mu_{i}}^{\rho}
{\delta {\cal T}_{\mu_{1},...,\hat{\mu_{i}},...,\mu_{n};
\nu_{1},...,\hat{\nu_{j}},...,\nu_{n}}\over\delta x^{\nu_{j}}} +
\delta_{\nu_{j}}^{\rho}
{\delta {\cal T}_{\mu_{1},...,\hat{\mu_{i}},...,\mu_{n};
\nu_{1},...,\hat{\nu_{i}},...,\nu_{n}}\over\delta
x^{\mu_{i}}}\right)\nonumber\\
{}~
\label{ADK_2}
\end{eqnarray}

So we have:
\begin{thm}
The most general local variational differential equation of
second order for a scalar field is given by (\ref{polyn}) where
the functions
$
{\cal T}_{...}
$
have the structure decribed in the statement of Theorem 2 and
also verify (\ref{ADK_1}) and (\ref{ADK_2}).
\label{structure'}
\end{thm}

3.6 We concentrate now on the form of possible Lagrangians
producing second order differential equations. According to
subsection 3.3 such a Lagrangian can be taken to be of second
order and constrained by (\ref{constr}). According to Theorem
\ref{structure}
this means that
${\cal L}$
can be taken of the form:
\be
{\cal L} = \sum_{k=0}^{n} {1\over (k!)^{2}}
{\cal L}_{\mu_{1},...,\mu_{k};\nu_{1},...,\nu_{k}}
\psi^{\mu_{1},...,\mu_{k};\nu_{1},...,\nu_{k}}
\label{lpolyn}
\ee
with
$
{\cal L}_{...}
$
independent of
$
\psi_{\{\rho\sigma\}}:
$
\be
\partial^{\rho\sigma} {\cal
L}_{\mu_{1},...,\mu_{k};\nu_{1},...,\nu_{k}} = 0~(k = 0,...,n)
\ee
and with the same symmetry properties as
$
{\cal T}_{...}.
$

Of course, it is possible that two different Lagrangians of the
type (\ref{lpolyn}) give the same Euler-Lagrange operator. To
investigate the extent of this arbitrariness we compute
$
E(L).
$
As expected, we get something of the form (\ref{polyn}):
\be
{\cal E}(L) = \sum_{k=0}^{n} {1\over (k!)^{2}}
{\cal T}(L)_{\mu_{1},...,\mu_{k};\nu_{1},...,\nu_{k}}
\psi^{\mu_{1},...,\mu_{k};\nu_{1},...,\nu_{k}}
\ee
where:
\begin{eqnarray}
&~&{\cal T}(L)_{\mu_{1},...,\mu_{k};\nu_{1},...,\nu_{k}} =
(n-k+1) \partial{\cal L}_{\mu_{1},...,\mu_{k};\nu_{1},...,\nu_{k}}
+ {\delta \over\delta x^{\rho}}
\left(\partial^{\rho} {\cal L}_{\mu_{1},...,\mu_{k};\nu_{1},...,\nu_{k}}
\right) + \nonumber
\\
&~&\sum_{i=1}^{k} (-1)^{i}
\left[ {\delta\over \delta x^{\mu_{i}}} \left(\partial^{\lambda}
{\cal L}_{\lambda,\mu_{1},...,\hat{\mu_{i}},...,\mu_{k};
\nu_{1},...,\nu_{k}}\right) + {\delta\over \delta x^{\nu_{i}}}
\left(\partial^{\lambda}
{\cal L}_{\mu_{1},...,\mu_{k};\lambda\nu_{1},...,
\hat{\nu_{i}},...,\nu_{k}}\right) \right] + \nonumber
\\
&~&\sum_{i,j=1}^{k} (-1)^{i+j}
{\delta^{2} {\cal L}_{\mu_{1},...,\hat{\mu_{i}},...,\mu_{k};
\nu_{1},...,\hat{\nu_{j}},...,\nu_{k}}\over
\delta x^{\mu_{i}}\delta x^{\nu_{j}}} -
\partial^{\lambda} \partial^{\zeta}
{\cal L}_{\lambda,\mu_{1},...,\mu_{k};\zeta,\nu_{1},...,\nu_{k}}
\end{eqnarray}
\hfill{(k = 0,...,n-2),}
\begin{eqnarray}
&~&{\cal T}(L)_{\mu_{1},...,\mu_{n-1};\nu_{1},...,\nu_{n-1}} =
{}~2 \partial{\cal L}_{\mu_{1},...,\mu_{n-1};\nu_{1},...,\nu_{n-1}}
+ {\delta \over\delta x^{\rho}} \left(\partial^{\rho}
{\cal L}_{\mu_{1},...,\mu_{n-1};\nu_{1},...,\nu_{n-1}}\right) +
\nonumber\\
&~&\sum_{i=1}^{n-1} (-1)^{i}
\left[ {\delta\over \delta x^{\mu_{i}}} \left(\partial^{\lambda}
{\cal L}_{\lambda,\mu_{1},...,\hat{\mu_{i}},...,\mu_{n-1};
\nu_{1},...,\nu_{n-1}}\right) +
{\delta\over \delta x^{\nu_{i}}}
\left(\partial^{\lambda}
{\cal L}_{\mu_{1},...,\mu_{n-1};\lambda,\nu_{1},...,
\hat{\nu_{i}},...,\nu_{n-1}}\right) \right]
\nonumber \\
&~&+ \sum_{i,j=1}^{n-1} (-1)^{i+j}
{\delta^{2} {\cal L}_{\mu_{1},...,\hat{\mu_{i}},...,\mu_{n-1};
\nu_{1},...,\hat{\nu_{j}},...,\nu_{n-1}}\over
\delta x^{\mu_{i}}\delta x^{\nu_{j}}}
\end{eqnarray}
\begin{eqnarray}
&~&{\cal T}(L)_{\mu_{1},...,\mu_{n};\nu_{1},...,\nu_{n}} =
\partial{\cal L}_{\mu_{1},...,\mu_{n};\nu_{1},...,\nu_{n}}
- {\delta \over\delta x^{\rho}}
\left(\partial^{\rho} {\cal L}_{\mu_{1},...,\mu_{n};\nu_{1},...,\nu_{n}}
\right) + \nonumber
\\
&~&\sum_{i,j=1}^{n} (-1)^{i+j}
{\delta^{2} {\cal L}_{\mu_{1},...,\hat{\mu_{i}},...,\mu_{n};
\nu_{1},...,\hat{\nu_{j}},...,\nu_{n}}\over
\delta x^{\mu_{i}}\delta x^{\nu_{j}}}
\end{eqnarray}

We use in these equations the Bourbaki convention
$
\sum_{\emptyset}\cdots = 0.
$
So,
${\cal L}$
given by (\ref{lpolyn}) leads to trivial Euler-Lagrange
equations {\it iff} the expressions
$
{\cal T}(L)_{...}
$
defined above are identically zero.

3.7 We are prepared to investigate now the most general
expression of a symmetry for a second order local variational
differential equation for a scalar field. We have:
\begin{thm}
Let
$T$
a local variational differential equation for a scalar field and
$
\phi \in Diff(S)
$
a symmetry. Then there exists
$
\rho \in {\cal F}(J_{n}^{s}(S))
$
such that:
\be
\partial^{\mu\nu} \rho = 0
\label{ind}
\ee
and
\be
(\dot\phi)^{*}~T = \rho T.
\label{sym3}
\ee
\label{symmetry}
\end{thm}

{\bf Proof}

The condition that
$
\phi
$
is a symmetry is that (\ref{ELeq}) should be equivalent to the
same equation with
$
\Psi \mapsto \phi\circ\Psi
$
for any evolution
$
\Psi:M \mapsto S.
$
Because
$\Psi$
is arbitrary one obtains that:
$$
T = 0 \Leftrightarrow (\dot\phi)^{*}~T = 0.
$$

Equivalenty, if we define
$
{\cal T}'
$
by
$$
(\dot\phi)^{*}~T = {\cal T}'~d\psi\wedge dx^{1}\wedge ...\wedge dx^{n}
$$
then we have:
$$
{\cal T} = 0 \Leftrightarrow {\cal T}' = 0.
$$

One easily obtains from here, under some reasonable regularity
conditions, that there exists a funtion
$
\rho \in {\cal F}(J_{n}^{s}(S))
$
such that
\be
{\cal T}' = \rho~{\cal T}.
\label{sym4}
\ee

Because
$T$
and
$T'$
are locally variational
$
{\cal T}
$
and
$
{\cal T}'
$
have the polynomial structure given by (\ref{polyn}). So we have
$
\rho = {p \over p'},
$
where
$p$
and
$p'$
are some polynomials in
$
\psi_{\{\mu\nu\}}.
$
So, (\ref{sym4}) is:
\be
p~{\cal T} = p'~{\cal T}'.
\label{sym5}
\ee

We identify the terms of maximal degree in
$
\psi_{\{\mu\nu\}}
$
in both sides and find
\be
p_{max}~{\cal T}_{\emptyset;\emptyset}~det(\psi) =
p'_{max}~ {\cal T}'_{\emptyset;\emptyset}~det(\psi)~\Leftrightarrow
p_{max} = \rho_{0}~p'_{max}
\ee
where
$
\rho_{0} \equiv {{\cal T}_{\emptyset;\emptyset} \over
{\cal T}'_{\emptyset;\emptyset}}.
$
We insert this in (\ref{sym5}) and continue by recurrence.
Finally one gets
$
\rho = \rho_{0}
$
so we have in fact (\ref{ind}). Moreover, it is clear that
(\ref{sym4}) is equivalent to (\ref{sym3}).
$\Box$

\begin{rem}
One can obtain some usefull relations from (\ref{sym4}) if we
insert it into (\ref{ADK_1}) and (\ref{ADK_2}) and take into
account that
${\cal T}$
verify these equations also. One obtains:
\begin{eqnarray}
&~&\sum_{i,j=1}^{k} (-1)^{i+j}
\left( \delta_{\mu_{i}}^{\rho} {\delta f\over\delta x^{\nu_{j}}} +
\delta_{\nu_{j}}^{\rho} {\delta f \over\delta x^{\mu_{i}}}\right)
{\cal T}_{\mu_{1},...,\hat{\mu_{i}},...,\mu_{k};
\nu_{1},...,\hat{\nu_{j}},...,\nu_{k}} \nonumber
\\
&~&= \sum_{i=1}^{k} (-1)^{i-1} \left( \delta_{\mu_{i}}^{\rho}
{\cal T}_{\lambda,\mu_{1},...,\hat{\mu_{i}},...,\mu_{k};
\nu_{1},...,\nu_{k}} + \delta_{\nu_{i}}^{\rho}
{\cal T}_{\mu_{1},...,\mu_{k};
\lambda,\nu_{1},...,\hat{\nu_{i}},...,\nu_{k}}\right)
{\partial f\over \partial \psi_{\lambda}}
\end{eqnarray}
\hfill{(k = 1,...,n-1),}
\be
{\partial f\over\partial\psi^{\rho}}
{\cal T}_{\mu_{1},...,\mu_{n};\nu_{1},...,\nu_{n}} =
{1\over2} \sum_{i,j=1}^{n} (-1)^{i+j}
\left( \delta_{\mu_{i}}^{\rho} {\delta f\over\delta x^{\nu_{j}}} +
\delta_{\nu_{j}}^{\rho} {\delta f\over\delta x^{\mu_{i}}}\right)
{\cal T}_{\mu_{1},...,\hat{\mu_{i}},...,\mu_{n};
\nu_{1},...,\hat{\nu_{j}},...,\nu_{n}}
\ee

These relations can be used to obtain some restrictions on the
function
$f$.
For instance, let us suppose that
$
{\delta f\over\delta x^{\lambda}} = 0
$
and
$
{\partial f\over\partial\psi^{\rho}} = 0.
$
Then one obtains that either
$
{\partial f\over\partial\psi} = 0
$
(in this case
$f$
is locally constant) or
$
{\cal T}_{...}
$
verifies:
\be
\sum_{i,j=1}^{k} (-1)^{i+j} \left( \delta_{\mu_{i}}^{\rho}
\psi_{\nu_{j}} + \delta_{\nu_{j}}^{\rho} \psi_{\mu_{i}}\right)
{\cal T}_{\mu_{1},...,\hat{\mu_{i}},...,\mu_{k};
\nu_{1},...,\hat{\nu_{j}},...,\nu_{k}} = 0~
(k = 1,...,n).
\label{LHC3}
\ee
\end{rem}

\begin{rem}
Theorem 2 is a sort of Lee-Hwa Chung theorem \cite{SM} for the
Lagrangian formalism.
\end{rem}

\section{Lagrangian Systems with Groups of Symmetries}

4.1 We will study two types of symmetry in this section. First,
the case when the group of symmetries is a Lie group (with a
typical case the Poincar\'e invariance) and next the case when the
group of symmetries is infinite dimensional (with the typical
case the universal invariance).

4.2 Let us consider a second order locally variational equation
with Poincar\'e invariance. (When speaking of Poincar\'e
invariance we will have in mind the proper orthochronous
Poincar\'e group, although there is no dificulty in treating the
inversions with the same method.)

So,
$M$
from 3.2 is the
$n$-dimensional
Minkowski space and for obvious reasons the indices
$
\mu, \nu,...
$
will take the values
$
0,1,...,n-1;
$
the Minkowski bilinear form
$G_{..}$
has the signature
$
(1,-1,...,-1).
$
The action of the Poincar\'e group on
$
S \equiv M \times \R
$
is
\be
\phi_{L,a}(x,\psi) = (Lx + a,\psi)
\label{p1}
\ee
with
$L$
a Lorentz transformation and
$
a \in \R^{n}
$
a translation in the affine space
$M$.
The lift of (\ref{p1}) to
$
J_{n}^{2}(S)
$
is:
\be
\dot\phi_{L,a}(x,\psi,\psi_{\mu},\psi_{\{\mu\nu\}}) =
(Lx + a,\psi,{L_{\mu}}^{\nu}\psi_{\nu},
{L_{\mu}}^{\rho}{L_{\nu}}^{\sigma}\psi_{\{\rho\sigma\}})
\ee
and the condition of Poincar\'e invariance is by definition:
\be
(\dot\phi_{L,a})^{*}~T = T
\label{PI1}
\ee
(so we are considering only Noetherian symmetries).

The equation (\ref{PI1}) is equivalent to:
\be
{\cal T}\circ \dot\phi_{L,a} = {\cal T}.
\label{PI2}
\ee

For
$
L = {\bf 1}
$
one obtains the
$x$-independence
of
${\cal T}$:
\be
{\partial {\cal T}\over \partial x^{\mu}} = 0
\label{PI3}
\ee
and from (\ref{PI2}) we still have the Lorentz invariance of
${\cal T}$:
\be
{\cal T}({\psi,L_{\mu}}^{\rho}{L_{\nu}}^{\sigma}\psi_{\{\rho\sigma\}}) =
{\cal T}(\psi,\psi_{\mu},\psi_{\{\mu\nu\}})
\label{PI4}
\ee

If we insert (\ref{polyn}) into (\ref{PI3}) and (\ref{PI4}) we
get that
$
{\cal T}_{...}
$
are
$x$-independent:
\be
{\partial {\cal T}_{\mu_{1},...,\mu_{k};\nu_{1},...,\nu_{k}}
\over \partial x^{\lambda}} = 0~(k = 0,...,n)
\ee
and also that
$
{\cal T}_{...}
$
are Lorentz covariant tensors depending only of
$\psi$
and
$
\psi_{\mu}.
$

Using the usual method \cite{D} of analysing the generic form of
such a tensorial covariant functions one obtains that
$
{\cal T}_{...}
$
is a sum of expressions of the type:
$$
\psi_{.}... \psi_{.} G_{..}...G_{..} {\cal A}(\psi,J)
$$
where
$
J \equiv \psi^{\mu}\psi_{\mu}
$
is a Lorentz invariant.

One has to take into account now the various symmetry properties
of
$
{\cal T}_{...}.
$
First one notices that one cannot have more than two factors
$
\psi_{.}
$
because for three factors or more one contradicts the
antisymmetry in
$
\mu_{1},...,\mu_{k}
$
or/and in
$
\nu_{1},...,\nu_{k}.
$
Because we also have symmetry with respect to the change
$
(\mu_{1},...,\mu_{k}) \leftrightarrow (\nu_{1},...,\nu_{k})
$
it is clear that we have two types of terms: terms containing no
$
\psi_{.}
$
factors and terms containing exactly two
$
\psi_{.}
$
factors, more precisely of the form
$
\psi_{\mu_{k}} \psi_{\nu_{l}}.
$
Also, to avoid contradicting of the antisymmetry the factors
$
G_{..}
$
alloweded are of the form
$
G_{\mu_{k}\nu_{l}}.
$

Summing up, the most general Lorentz covariant tensor
$
{\cal T}_{...}
$
respecting the symmetry properties from the statement of Theorem
2 is
\be
{\cal T}_{\mu_{1},...,\mu_{k};\nu_{1},...,\nu_{k}} = {\cal A}_{k}
I_{\mu_{1},...,\mu_{k};\nu_{1},...,\nu_{k}} + {\cal B}_{k}
J_{\mu_{1},...,\mu_{k};\nu_{1},...,\nu_{k}}.
\label{INV2}
\ee

Here
$
{\cal A}_{k}
$
and
$
{\cal B}_{k}
$
are smooth functions of
$\psi$
and
$J$.
We use the convention
$
{\cal B}_{0} = 0
$
and we have defined
\be
I_{\mu_{1},...,\mu_{k};\nu_{1},...,\nu_{k}} =
\sum_{\sigma,\tau \in {\cal P}_{\{1,...,k\}}} (-1)^{\vert\sigma\vert+
\vert\tau\vert} \prod_{i=1}^{k} G_{\mu_{\sigma(i)}\nu_{\tau(i)}}~
(k = 0,...,n)
\ee
\be
J_{\mu_{1},...,\mu_{k};\nu_{1},...,\nu_{k}} =
\sum_{\sigma,\tau \in {\cal P}_{\{1,...,k\}}} (-1)^{\vert\sigma\vert+
\vert\tau\vert} \psi_{\mu_{\sigma(1)}}\psi_{\nu_{\tau(1)}}
\prod_{i=2}^{k} G_{\mu_{\sigma(i)}\nu_{\tau(i)}}~
(k = 0,...,n)
\ee
with the conventions
$
I_{\emptyset;\emptyset} \equiv 1,~J_{\emptyset;\emptyset} \equiv 0.
$

One must insert (\ref{INV2}) into the remaining ADK equations
(\ref{ADK_1}) and (\ref{ADK_2}). The result of this tedious
computation is:
\be
{\partial {\cal A}_{k-1}\over \partial \psi} -
2k {\partial {\cal A}_{k}\over \partial J} -
2J {\partial {\cal B}_{k}\over \partial J} - (n + k) {\cal
B}_{k} = 0~(k = 1,...,n-1)
\label{einv1}
\ee
\be
{\partial {\cal B}_{k}\over \partial \psi} = 0~(k = 1,...,n-2)
\label{einv2}
\ee
\be
2 {\partial {\cal A}_{n}\over \partial J} +
(n-1)! {\partial {\cal A}_{n-1}\over \partial \psi} = 0
\label{einv3}
\ee
where we understand that for
$
n = 2,
$
(\ref{einv2}) dissapears. Inserting (\ref{INV2}) into
(\ref{polyn}) it follows that we have:
\begin{thm}
The most general local variational differential equation of
second order for a scalar field having Poincar\'e invariance in
the sense (\ref{PI2}) is of the form:
\be
{\cal T} = {\cal A}_{0} det(\psi) + \sum_{k=1}^{n-1}
\left( {\cal A}_{k} I_{k} + {\cal B}_{k} J_{k} \right) + {\cal A}_{n}
\label{INV3}
\ee
where
$
I_{k}
$
and
$
J_{k}
$
are the Lorentz invariants:
\be
I_{k} \equiv \left(\prod_{i=1}^{k} G_{\mu_{i}\nu_{i}}\right)
\psi^{\mu_{1},...,\mu_{k}; \nu_{1},...,\nu_{k}}
\ee
and
\be
J_{k} \equiv \psi_{\mu_{1}} \psi_{\nu_{1}}
\left(\prod_{i=2}^{k} G_{\mu_{i}\nu_{i}}\right)
\psi^{\mu_{1},...,\mu_{k}; \nu_{1},...,\nu_{k}}.
\ee

Also the functions
$
{\cal A}_{0},...,{\cal A}_{n}
$
and
$
{\cal B}_{1},...,{\cal B}_{n-1}
$
depend smoothly only of
$
\psi
$
and
$J$
and verify the equations (\ref{einv1})-(\ref{einv3}). One can
take
$
{\cal B}_{1},...,{\cal B}_{n-2}
$
arbitrary functions of
$J$
and
$
{\cal A}_{n},{\cal B}_{n-1}
$
arbitrary functions of
$
\psi
$
and
$J$.
Then (\ref{einv1})-(\ref{einv3}) can be used to fix
$
{\cal A}_{0},...,{\cal A}_{n-1}
$
up to an arbitrary function of
$J$.
The Tonti Lagrangian has the structure (\ref{INV3}) also.
\label{Poincare}
\end{thm}

4.3 Let us study now the so-called universal invariance. Suppose
$
F \in Diff(\R);
$
then we define
$
\phi_{F} \in Diff(S)
$
by:
\be
\phi_{F}(x,\psi) = (x,F(\psi)).
\ee

The natural lift of
$
\phi_{F} \in Diff(S)
$
to
$
J_{n}^{2}(S)
$
is:
\be
\dot\phi_{F}(x,\psi,\psi_{\mu},\psi_{\{\mu\nu\}}) =
(x,F(\psi),F'(\psi)\psi_{\mu},F'(\psi)\psi_{\{\mu\nu\}}+
F''(\psi)\psi_{\mu}\psi_{\nu}).
\ee

We say that the differential equation
$T$
has {\it universal invariance} if we have:
\be
(\dot\phi_{F})^{*}~T = \rho_{F}~T
\label{UI1}
\ee

The function
$
\rho_{F} \in Diff(J_{n}^{2}(S))
$
does not depend on
$
\psi_{\{\mu\nu\}}
$
according to Theorem \ref{symmetry} and it is a cohomological object
\cite{G2}. As in \cite{G2} we will consider only the case when:
\be
\rho_{F} = (F')^{p}.
\ee

In this case (\ref{UI1}) is equivalent to:
\be
{\cal T}\circ\dot\phi_{F} = (F')^{p-1}~{\cal T}.
\label{UI3}
\ee

\begin{rem}
According to Remark 2, we have two cases: either
$
p = 1
$
or we have (\ref{LHC3}).
\end{rem}

We take
$F$
to be an infinitesimal diffeomorphism i.e.
\be
F(\psi) = \psi + \theta(\psi)
\ee
with
$
\theta
$
infinitesimal but otherwise arbitrary and we can cast
(\ref{UI3}) into the infinitesimal form; one obtains:
\be
\partial {\cal T} = 0
\label{EqINV1}
\ee
\be
\psi_{\mu} \partial^{\mu} {\cal T} +
\psi_{\mu}\psi_{\nu} \partial^{\mu\nu} {\cal T} =
(p - 1) {\cal T}
\label{EqINV2}
\ee
\be
\psi_{\mu} \psi_{\nu} \partial^{\mu\nu} {\cal T} = 0.
\label{EqINV3}
\ee

Let us note that (\ref{EqINV2}) is the infinitesimal form of the
homogeneity equation:
\be
{\cal T}(x,\lambda\psi_{\mu},\lambda\psi_{\{\mu\nu\}}) =
\lambda^{p-1}~{\cal T}(x,\psi_{\mu},\psi_{\{\mu\nu\}})~(\forall
\lambda \in \R^{*}).
\ee

If we insert in these equations the expression (\ref{polyn}) we
obtain equivalently:
\be
\partial {\cal T}_{\mu_{1},...,\mu_{k};\nu_{1},...,\nu_{k}}
= 0~(k = 0,...,n)
\label{EqINV_1}
\ee
\be
{\cal T}_{\mu_{1},...,\mu_{k};\nu_{1},...,\nu_{k}}
(x,\lambda\psi_{\mu}) = \lambda^{k+p-1-n}
{\cal T}_{\mu_{1},...,\mu_{k};\nu_{1},...,\nu_{k}}(x,\psi_{\mu})~
(k = 0,...,n)
\label{EqINV_2}
\ee
\be
\sum_{i,j=1}^{n} (-1)^{i+j} \psi_{\mu_{i}} \psi_{\nu_{j}}
{\cal T}_{\mu_{1},...,\hat{\mu_{i}},...,\mu_{k};
\nu_{1},...,\hat{\nu_{j}},...,\nu_{k}} = 0.~
(k = 1,...,n)
\label{EqINV_3}
\ee

Let us note that for
$
p \not= 0,
$
(\ref{EqINV_3}) follows from (\ref{LHC3}).

One must add to these equations (\ref{ADK_1}) and (\ref{ADK_2})
which are in our case:
\begin{eqnarray}
&~&\sum_{i,j=1}^{k} (-1)^{i+j} \left( \delta_{\mu_{i}}^{\rho}
{\partial {\cal T}_{\mu_{1},...,\hat{\mu_{i}},...,\mu_{k};
\nu_{1},...,\hat{\nu_{j}},...,\nu_{k}}\over\partial x^{\nu_{j}}} +
\delta_{\nu_{j}}^{\rho}
{\partial {\cal T}_{\mu_{1},...,\hat{\mu_{i}},...,\mu_{k};
\nu_{1},...,\hat{\nu_{j}},...,\nu_{k}}\over\partial
x^{\mu_{i}}}\right) = \nonumber
\\
&~&\sum_{i=1}^{k} (-1)^{i-1} \left[ \delta_{\mu_{i}}^{\rho}
\left(\partial^{\lambda}
{\cal T}_{\lambda,\mu_{1},...,\hat{\mu_{i}},...,\mu_{k};
\nu_{1},...,\nu_{k}}\right) + \delta_{\nu_{i}}^{\rho}
\left(\partial^{\lambda} {\cal
T}_{\mu_{1},...,\mu_{k};\lambda,\nu_{1},...,\hat{\nu_{i}},...,\nu_{k}}
\right) \right]
\label{EqINV_4}
\end{eqnarray}
\hfill{(k = 1,...,n-1),}
\begin{eqnarray}
&~&\partial^{\rho}
{\cal T}_{\mu_{1},...,\mu_{n};\nu_{1},...,\nu_{n}} =\nonumber
\\
&~&{1\over2} \sum_{i,j=1}^{n} (-1)^{i+j} \left( \delta_{\mu_{i}}^{\rho}
{\partial {\cal T}_{\mu_{1},...,\hat{\mu_{i}},...,\mu_{n};
\nu_{1},...,\hat{\nu_{j}},...,\nu_{n}}\over\partial x^{\nu_{j}}} +
\delta_{\nu_{j}}^{\rho}
{\partial {\cal T}_{\mu_{1},...,\hat{\mu_{i}},...,\mu_{n};
\nu_{1},...,\hat{\nu_{j}},...,\nu_{n}}\over\partial
x^{\mu_{i}}}\right)\nonumber\\
{}~
\label{EqINV_5}
\end{eqnarray}

The system (\ref{EqINV_1})-(\ref{EqINV_5}) seems to be too hard
to solve in the general case. We content ourselves to study two
particular cases.

(a)
$T$
is translational invariant i.e.
\be
{\partial {\cal T} \over\partial x^{\lambda}} = 0
\ee
or:
\be
{\partial {\cal T}_{\mu_{1},...,\mu_{k};\nu_{1},...,\nu_{k}}
\over\partial x^{\lambda}} = 0~(k = 0,...,n).
\ee

For
$
p = n + 1
$
one obtains the particular solution
\be
{\cal T} = {\cal T}_{\emptyset;\emptyset} det(\psi)
\ee
with
$
{\cal T}_{\emptyset;\emptyset}
$
constant. This is the solution appearing in \cite{F}.

(b) It is clear that
$T$
follows from a first order Lagrangian {\it iff}
\be
{\cal T}_{\mu_{1},...,\mu_{k};\nu_{1},...,\nu_{k}} = 0~(k = 0,...,n-2).
\ee

In this case:
\be
{\cal T} = {\cal T}_{0} + {\cal T}^{\rho\sigma} \psi_{\{\rho\sigma\}}.
\ee

One easily obtains that (\ref{EqINV_1})-(\ref{EqINV_5}) reduces to:
\be
\partial {\cal T}_{0} = 0
\ee
\be
\partial {\cal T}^{\rho\sigma} = 0
\ee
\be
{\cal T}_{0}(x,\lambda\psi_{\mu}) = \lambda^{p-1}
{\cal T}_{0}(x,\psi_{\mu})
\ee
\be
{\cal T}^{\rho\sigma}(x,\lambda\psi_{\mu}) = \lambda^{p-2}
{\cal T}^{\rho\sigma}(x,\psi_{\mu})
\ee
\be
\psi_{\rho}\psi_{\sigma} {\cal T}^{\rho\sigma} = 0
\ee
\be
\partial^{\rho} {\cal T}^{\mu\nu} - \partial^{\mu} {\cal
T}^{\rho\nu} = 0
\ee
\be
\partial^{\rho} {\cal T}_{0} = {\partial {\cal T}^{\rho\sigma}
\over \partial x^{\sigma}}.
\ee

This system was analysed in \cite{G2} where it was found that it
has solutions for
$
p = 0
$
and
$
p = 1.
$

\section{Conclusions}

The central formula obtained in this paper is (\ref{polyn}).
This expression affords a rather complete treatement of local
variational differential equations of second order with groups
of symmetry.

It is plausible that (\ref{polyn}) admits generalizations for
the case
$
N > 1
$
(i.e. fields with more than one components) and for
$
s > 2
$
(i.e. equations of arbitrary order). Maybe as a first step one
should try the more modest cases:
$
N > 1,~s = 2
$
or
$
N = 1,~s > 2.
$

These problems will be adressed in further publications.

\end{document}